\documentclass[aps,prb,twocolumn,superscriptaddress]{revtex4-2}
\usepackage{bm}
\usepackage{ae}
\usepackage[T1]{fontenc}
\usepackage[ansinew]{inputenc}
\usepackage{amsmath}
\usepackage{amssymb}
\usepackage{graphicx}
\usepackage{color}
\usepackage[colorlinks]{hyperref}
\usepackage{epstopdf}
\input{epsf}

\begin{document}

\title{Creation of two-dimensional circular motion of charge qubit}

\author{L.Y. Gorelik}
\affiliation{Department of Physics, Chalmers University of Technology, SE-412 96 G{\"o}teborg, Sweden}

\author{S.I. Kulinich}
\affiliation{B. Verkin Institute for Low Temperature Physics and Engineering of the National Academy of Sciences of Ukraine, 47 Prospekt Nauky, Kharkiv 61103, Ukraine}

\author{R.I. Shekhter}
\affiliation{Department of Physics, University of Gothenburg, SE-412 96 G\"oteborg, Sweden}

\author{D. Radi\'{c}*}
\affiliation{Department of Physics, Faculty of Science, University of Zagreb, Bijeni\v{c}ka 32, Zagreb 10000, Croatia}

%\date{\today}
%\pacs{}

\begin{abstract}
We suggest a nanoelectromechanical setup which generates a particular type of motion - the circular motion of mesoscopic superconducting grain, where motion is described by entangled nanomechanical coherent states. The setup is based on mesoscopic terminal utilizing the AC Josephson effect between the superconducting electrodes and the grain, operating in the regime of the Cooper pair box controlled by the gate voltage. The grain is placed on the free end of the suspended cantilever, performing controlled two-dimensional mechanical vibrations. Required functionality is achieved by operating two external parameters, bias voltage between the superconducting electrodes and voltage between gate electrodes, by which the nanomechanical coherent states are formed and organised in a pair of entangled cat-states in two perpendicular spatial directions which evolve in time in the way to provide a circular motion.   
\end{abstract}

\maketitle

\section{Introduction}

Qubit \cite{Schumacher} is a basic constituent of any device capable of containing or processing quantum information \cite{Nielsen}. Beside the problem of storing and perform computing with quantum information, another formidable problem is its physical transportation between different processing terminals in space.
The implementation of qubit has been subject of many fields of physics, the most popular among them being from optics in resonant cavities, cold atoms, to the solid state implementations as magnetic fluxons in Josephson junctions \cite{QuantumProcessing,Girvin,Devoret,Mirhosseini,Leibfried} in an attempt to find an optimum between cons and pros among them.  
In spite of significantly higher level of development and application of the above-mentioned implementations, in the ongoing line of research, which this paper is part of, we focus on nanoelectromechanical (NEM) implementations of qubit \cite{npjDR,Bahrova} and processing of quantum information. In that respect we entangle the state of charge qubit\cite{Bouchiat,Nakamura,Robert,Lehnert} with its nanomechanical vibrational excitations \cite{Schneider,Hann,Chu} in the form of coherent states.
Choice of the NEM implementation brings forward a unique level of compactness, with easily electrically controlled charge-qubit and nanomechanics with amazingly high quality factors achievable nowadays \cite{Laird,Tao,Bereyhi}.
So far, we have suggested the mechanism for quantum information transferring between terminals within the quantum network \cite{PhysB_DR2}, utilizing the means of its transduction \cite{PhysB_DR} between electrical and mechanical vibrational degrees of freedom of physically coupled charge qubits to mechanical resonator. 
Another open question is if it is possible to achieve transferring of quantum information by designing a spatial motion of a qubit between nodes within the quantum network, using the, so-called, concept of "flying qubits".
In order to achieve such a goal, the first step would be finding the means of engineering the desired spatial trajectory of the physical object constituting the qubit. In this paper we suggest a NEM setup in which, by operating the controlling parameters (i.e. external voltages on electrodes) we achieve a two-dimensional circular motion of the superconducting grain in the regime of the Cooper pair box qubit.

\section{The Model}

\begin{figure}
\centerline{\includegraphics[width=.9\columnwidth]{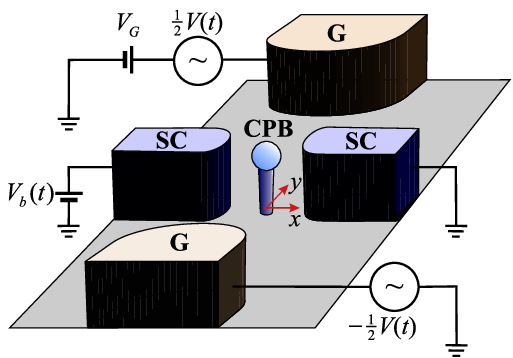}}
\caption{Schematic illustration of the NEM setup. Two superconducting electrodes (SC) are voltage-biased by $V_b(t)$, controllable over time $t$. Between them there is a cantilever with a superconducting mesoscopic grain (CPB) on the top, capable of performing the 2D mechanical vibrations in the $(x,y)$-plane. Position-dependent tunneling of Cooper pairs takes place between the CPB and SC electrodes. The role of gate electrodes (G) is twofold: (1) maintaining the state of the superconducting grain in the regime of the Cooper pair box (CPB) by $V_G$; (2) providing the time-dependent AC electric field by $V(t)$, to manipulate the states of the CPB in the desired way.}
\label{Fig-Schematics}
\end{figure}

The suggested NEM is a mesoscopic terminal based on the AC Josephson effect (see Fig. \ref{Fig-Schematics}). It contains two voltage-biased superconducting electrodes (SC), with a superconducting mesoscopic grain, placed on a top of a nanomechanical beam between them. States in the grain are controlled by the gate (G) voltage in the twofold way: (1) By constant voltage $V_G$, states in the superconducting grain are tuned to be in regime of the Cooper pair box (CPB). (2) In addition, there is an high-frequency oscillating voltage $V(t)$ by which we manipulate the states in the desired way (still preserving the CPB regime, i.e. perturbation is small comparing to the superconducting gap) and control functionality of the device. 
The nanomechanical beam is suspended on one end, performing the in-plane oscillations in two orthogonal directions approximately within the $(x,y)$-plane.
%
%
%%%%%%%%%%%%%%%%%%%%%%%%%%
%%%%%%%%%%%%%%%%%%%%%%%%%%%%
%
%
The superconducting part of the junction (SC electrodes and CPB oscillating between them) operates, as mentioned, in the regime of the AC Josephson effect, i.e. the superconducting electrodes are biased by a constant symmetric bias voltage $V_b(t)$, which can be switched on to value $V_b$ or off to zero, providing a superconducting phase difference $\Phi(t)$ dependent on time $t$. Here, $\Phi(t)=\mathrm{sgn}(V_b)\,\Omega t$ and $\Omega = 2|e V_b|/\hbar$ is the Josephson frequency (some constant initial phase at $t=0$ is for simplicity taken to be zero), $e$ is the electron charge. Cooper pairs tunnel between the SC-electrodes and the CPB oscillating on the cantilever (beam) at frequency $\omega$ assumed to be equal in both $x$- and $y$-direction. It is important to stress that the tunneling is essentially position-dependent. Neglecting a geometric asymmetry of the junction, we expand the Josephson coupling in terms of a small parameter $\varepsilon \equiv x_0 / x_{\text{tun}} \ll 1$, where $x_0=\sqrt{\hbar /m\omega}$ is the amplitude of zero-mode oscillations ($m$ is mass of the oscillator) and $x_{\text{tun}}$ is the characteristic tunneling length.\\

The CPB regime in the SC grain refers to the effective two-level system of degenerate states with zero and one excess Cooper pair, $\mathbf{e}_2^{\mp}$ respectively.
Those we call "the charge qubit states" or "the CPB qubit states". 
Additional function of symmetrically placed gate electrodes is to create approximately homogeneous corresponding electric field $\mathcal{E}(t)$ by applying an additional voltage $V(t)$ over the electrodes. This field appears along the central part of the junction where the CPB moves, with a zero-value of the corresponding electrostatic potential in the middle. 
Switching the $V(t)$ on and off, as well as $V_b$, is an important part of time-protocols of achieving the desired functionality as will be shown in the following section.\\

We write the time-dependent Hamiltonian describing the system, with couplings linear in $x$- and $y$-displacement of the CPB, in the form $H(t)=H_0(t)+H_{int}(t)$, where
% 
%%%%%%%%%%%%%%%%%%%%%%%%%% 
\begin{eqnarray}\label{Hamiltonian}
H_0(t)&=&E_J\sigma_1 \cos\Phi(t)+\hbar\omega(a^\dag a+ b^\dag b),\nonumber \\
H_{\text{int}}(t)&=&\epsilon_x (a+a^\dag)\sigma_2 \sin\Phi(t)+\epsilon_y (t) (b+b^\dag)\,\sigma_3.
\end{eqnarray}
%%%%%%%%%%%%%%%%%%%%%%%%%%
%
In Eq.(\ref{Hamiltonian}), $H_0(t)$ represents the noninteracting part of total Hamiltonian $H(t)$, comprising the position-independent part of Josephson tunneling term, characterised by the Josephson energy $E_J$, and 2D harmonic oscillator characterised by frequency $\omega$. Operators $\sigma_{1,2,3}$ are the Pauli matrices operating in the  $2 \times 2$ Hilbert subspace of the CPB, while  $a^\dag$ and $b^\dag$ are phonon creation operators for $x$- and $y$-direction of oscillator motion respectively. $H_{\text{int}}(t)$ is the interacting part of total Hamiltonian, describing position-dependent term of the Josephson coupling and coupling of the CPB to gate electrodes, both linear in displacements described by operators $\hat X= (a+a^\dag)/\sqrt{2}$ and $\hat Y= (b+b^\dag)/\sqrt{2}$ respectively. Josephson coupling is characterised by $\epsilon_x \equiv \varepsilon E_J$. Coupling to the gate is $\epsilon_y(t) \equiv 2|e|\mathcal{E}(t)x_0$, where $\mathcal{E}(t)$ is electric field created by voltage $V (t)$ on gate electrodes. Depending on demand, we can operate this electric field both as a constant signal, or signal periodic in time of magnitude $\mathcal{E}_0$ corresponding to the coupling constant $\epsilon_y = 2|e|\mathcal{E}_0 x_0$.
In our approach we assume that both couplings are weak, i.e. $\epsilon_x \ll \hbar \omega$ and $\epsilon_y \ll \hbar \omega$.

\section{The operating time-evolution protocols}

The desired functionality of the device is achieved by operating the control parameters, i.e. $V_b(t)$ and $V(t)$, along specific intervals of time, turning them on/off, correspondingly designing the time-evolution of the quantum system from the initial state. Those we call time-evolution protocols. Each protocol corresponds to the specific quantum time-evolution operator, governed by the Hamiltonian, which will evolve the system into specific quantum state described by the wave function $\mathbf \Psi(t)$, starting with the given initial state of the system $\mathbf \Psi(t=0)$. We assume that, initially, state of the system is prepared as
% 
%%%%%%%%%%%%%%%%%%%%%%%%%% 
\begin{eqnarray}\label{Psi_0}
\mathbf \Psi(0)= \mathbf e_{in} \otimes |0\rangle_x \otimes |0\rangle_y,
\end{eqnarray}
%%%%%%%%%%%%%%%%%%%%%%%%%%
%
where $\vert 0 \rangle_x \otimes |0\rangle_y$ is a zero-phonon ground state of mechanical subsystem (the 2D oscillator), while $\mathbf e_{in}$ is the initial charge-qubit state (where the quantum information can be encoded into superposition of the qubit's states to be "carried" further). We choose to work with projections to $\mathbf e_2^\pm$ vectors characterised by coefficients $c_\pm = \mathbf e_2^\pm \mathbf e_{in}$. Vectors $\mathbf{e}_i^{\pm}$, $i \in \{ 1,2,3 \}$ are the eigenvectors of $\sigma_i$ Pauli matrices corresponding to eigenvalues $\pm 1$.\\

The time-evolution operator $\hat U(t,t_0)$ of the wave function, from time moment $t_0$ to $t$, is generally a solution of the equation $i\hbar \tfrac{\partial}{\partial t}\hat U(t,t_0)=H(t) \hat U(t,t_0)$, with initial condition $\hat U(t_0,t_0)=\mathbf{1}$. Our goal is creating nanomechanical coherent states entangled to charge-qubit state. At $t=0$ we switch on the bias voltage from zero to constant value $V_b$ to provide finite phase difference $\Phi(t)$ in the SC electrodes. As shown in Refs. \cite{npjDR,Bahrova}, coupling of charge-qubit to mechanical subsystem under certain conditions leads to developing of mechanical coherent states. In that respect, in what follows, we restrict our consideration to the resonant case between Josephson and mechanical frequency, i.e. $\Omega=\omega$. The other condition we find to provide the desired functionality is setting the periodic signal $V(t)$ on the gate electrode, with frequency $2\omega$, i.e. to provide $\epsilon_y(t)=\epsilon_y \sin (2\omega t)$.
Under these conditions, the Hamiltonian $H(t)$, defined by Eq.(\ref{Hamiltonian}), is periodic in time with period $T=2\pi/\omega$.

We calculate the evolution operator $\hat U(t,t')$ within the interaction picture, i.e.
%
%%%%%%%%%%%%%%%%%%%%%%%%
\begin{eqnarray}\label{U1}
\hat U(t,t')=\hat u_0(t)\hat u(t,t')\hat u^\dag_0 (t'),
\end{eqnarray}
%%%%%%%%%%%%%%%%%%%%%%%%
%
with boundary condition $\hat u(t,t) = \mathbf 1$, where
%
%%%%%%%%%%%%%%%%%%%%%%%%
\begin{eqnarray}\label{u_0}
\hat u_0(t)= \exp\left[-i \omega t (a^\dag a + b^\dag b)- i \alpha\sigma_1 \sin \Phi(t) \right],
\end{eqnarray}
%%%%%%%%%%%%%%%%%%%%%%%%
%
and $\alpha \equiv E_J/\hbar\omega$ is the scale defined by the ratio of Josephson and mechanical energy.
Then, the equation for the evolution operator $\hat u(t,t')$ attains the form
%
%%%%%%%%%%%%%%%%%%%%%%%%
\begin{eqnarray}\label{EqOfMotion1}
i \hbar\frac{\partial \hat u(t,t')}{\partial
t}= H_{\text{eff}}(t) \hat u(t,t'),
\end{eqnarray}
%%%%%%%%%%%%%%%%%%%%%%%%
%
where $H_{\text{eff}}(t)=H^{(x)}_{\text{eff}}(t)+H^{(y)}_{\text{eff}}(t)$ is defined by
%
%%%%%%%%%%%%%%%%%%%%%%%%
\begin{eqnarray}\label{HeffXY}
H^{(x)}_{\text{eff}}(t)=\frac{\epsilon_x \sin \Phi(t)}{2}\left(a e^{-i\omega t}+a^\dag e^{i\omega t}\right) \nonumber\\
\times \left[\left(\sigma_2 +i\sigma_3 \right)e^{2i \alpha\sin\Phi(t)} + \text{h.c.} \right], \nonumber \\
H^{(y)}_{\text{eff}}(t)=\frac{\epsilon_y \sin2\omega
t}{2}\left(b e^{-i\omega t}+b^\dag e^{i\omega t}\right) \nonumber \\
\times \left[\left(\sigma_3 +i \sigma_2 \right)e^{2i \alpha\sin\Phi(t)} +\text{h.c.}\right].
\end{eqnarray}
%%%%%%%%%%%%%%%%%%%%%%%%
%
Under the resonance condition, $H_{\text{eff}}(t)$ is a periodic function with the same period $T$ as the original Hamiltonian $H(t)$. As a consequence, the evolution operator $\hat u(t,t')$ has a property
%
%%%%%%%%%%%%%%%%%%%%%%%%
\begin{eqnarray}\label{u-periodicity1}
\hat u(t-T,t'-T)=\hat u(t,t'),
\end{eqnarray}
%%%%%%%%%%%%%%%%%%%%%%%%
%
leading to
%
%%%%%%%%%%%%%%%%%%%%%%%%
\begin{eqnarray}\label{u-periodicity2}
\hat u(NT,0)= \hat u(NT,NT-T) \hat u(NT-T,NT-2T)... \nonumber\\
... \hat u(T,0)=\hat u^N(T,0), \,\,\,\,\,
\end{eqnarray}
%%%%%%%%%%%%%%%%%%%%%%%%
%
where $N$ is a positive integer number.

Using the equation of motion (\ref{EqOfMotion1}), we obtain the operator $\hat u(T,0)$ in the form
%
%%%%%%%%%%%%%%%%%%%%%%%%
\begin{eqnarray}\label{u(T,0)}
\hat u(T,0)=\mathbf 1 -\eta_x \sigma_2 \left(a-a^\dag\right)
- i \eta_y \sigma_2 \left(b+b^\dag\right)+{\cal{O}} (\epsilon_{x,y}^2), \,\,\,\,\,\,\,\,\,\,
\end{eqnarray}
%%%%%%%%%%%%%%%%%%%%%%%%
%
where
%
%%%%%%%%%%%%%%%%%%%%%%%%
\begin{eqnarray}\label{etaXY}
\eta_x=\frac{\pi\epsilon_x}{\hbar\omega}\left[J_0(2\alpha)-
J_2(2\alpha)\right], \nonumber\\
\eta_y=\frac{\pi\epsilon_y}{\hbar\omega}\left[J_1(2\alpha)+
J_3(2\alpha)\right],
\end{eqnarray}
%%%%%%%%%%%%%%%%%%%%%%%%
%
and $ J_n(2\alpha)$ is the Bessel function of the first kind.
Neglecting corrections ${\cal{O}} (\epsilon_{x,y}^2)$ of the order higher than linear in $\epsilon_{x,y}$, using Eqs.(\ref{U1},\ref{u-periodicity2},\ref{u(T,0)}), we obtain the evolution operator of the system for $t=NT$, i.e. for integer number of periods, in the form
%
%%%%%%%%%%%%%%%%%%%%%%%%
\begin{eqnarray}\label{U(NT,0)}
\hat U(NT,0) &=& e^{-2i\pi N(a^\dag a+b^\dag b)} \nonumber\\
&\times & \left[\mathbf 1 -\eta_x \sigma_2 \left(a-a^\dag\right) -i \eta_y \sigma_2 \left(b+b^\dag\right)\right]^N \nonumber\\
&\simeq & e^{-N\sigma_2 \left[\eta_x \left(a-a^\dag\right)+
i\eta_y \left(b+b^\dag\right)\right]} e^{-2i\pi N(a^\dag
a+b^\dag b)}.\,\,\,\,\,\,\, \nonumber\\
\end{eqnarray}
%%%%%%%%%%%%%%%%%%%%%%%%
%
The expression Eq.(\ref{U(NT,0)}) is valid under conditions
$N\epsilon_{x,y} \sim 1$ and $N\epsilon_{x,y}^2 \ll 1$.

\section{Results}

Accordingly to the Eq.(\ref{U(NT,0)}), the wave function $\mathbf \Psi (NT) = \hat U(NT,0) \mathbf \Psi (0)$ in the time moment $t=NT$ is obtained in the form
%
%%%%%%%%%%%%%%%%%%%%%%%%
\begin{equation}\label{Psi(NT)}
\mathbf \Psi(NT) = c_+\mathbf e_2^+ \otimes \vert Z_x\rangle\otimes \vert iZ_y \rangle + c_-\mathbf e_2^- \otimes \vert
-Z_x \rangle \otimes \vert -iZ_y \rangle,
\end{equation}
%%%%%%%%%%%%%%%%%%%%%%%%
%
where $\vert Z_x \rangle$ and $\vert Z_y \rangle$ are nanomechanical coherent states for $x$- and $y$-direction of oscillator motion, characterised by eigenvalues $Z_x= N \eta_x$ and $Z_y=-N \eta_y,$ being proportional to amplitudes, i.e. displacements from the origin of motion.

Using the result Eq.(\ref{Psi(NT)}), dividing the arbitrary time interval as $t=NT+\tau$, one can find the wave function at the arbitrary moment of time as $\mathbf \Psi(t) = \hat U(NT+\tau, NT) \mathbf \Psi(NT) \simeq \hat u_0(\tau) \mathbf
\Psi(NT)$, yielding the general result
%
%%%%%%%%%%%%%%%%%%%%%%%%
\begin{eqnarray}\label{Psi(t)}
\mathbf \Psi(t) &\simeq &  e^{-i \alpha \sigma_1 \sin \Phi(\tau)} 
\left[c_+\mathbf e_2^+ \otimes \vert Z_x(\tau)\rangle\otimes \vert iZ_y(\tau)\rangle \right. \nonumber\\
&+ & \left.  c_- \mathbf e_2^- \otimes \vert -Z_x(\tau) \rangle \otimes \vert -iZ_y(\tau) \rangle \right],
\end{eqnarray}
%%%%%%%%%%%%%%%%%%%%%%%%
%
where $Z_{x,y}(\tau)=Z_{x,y}\exp(-i\omega\tau)$ and $\tau = t-NT$, $N=\mathrm{floor}(t/T)$. Result is approximate in the sense that during the short time interval after the $N$-period evolution ($N \gg 1$), we neglected slow, proportional to small $\epsilon_{x,y}$, "growth" of coherent states. It would be exact if we switch off $\epsilon_x$ coupling (i.e. set $V_b=0$), after the time moment $t=NT$. Note that pair of coherent states $\vert \pm Z_x \rangle$, and pair of coherent states $\vert \pm iZ_y \rangle$ represent 2 independent degrees of freedom of mechanical motion along $x$- and $y$-direction.
In that sense, the wave function Eq.(\ref{Psi(t)}) represents the entangled state of three objects, $\mathbf e_2$, $\vert Z_x \rangle$ and $\vert iZ_y \rangle$, i.e. of charge-qubit and two nanomechanical cat states.\\

\begin{figure}
\centerline{\includegraphics[width=.75\columnwidth]{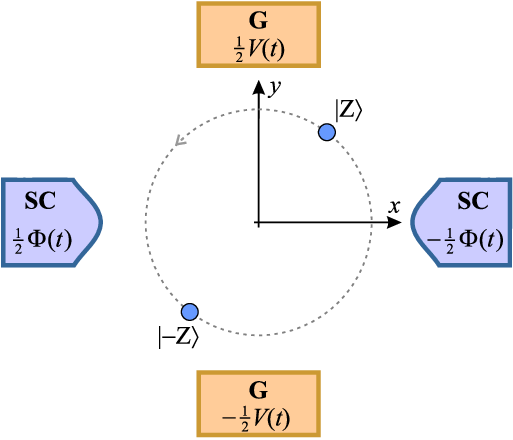}}
\caption{A pair of 2D nanomechanical coherent states is formed by the coherent dynamics of the NEM setup, $\mid \pm Z(x,y;t) \rangle \equiv \mid \pm Z_x(x,p_x;t) \rangle \, \otimes \mid \pm iZ_y(y,p_y;t) \rangle$, in the most general sense appearing as an entangled state of coherent states for motion in $x$- and $y$-direction (and also the CPB state $\mathbf{e}_2^{\pm}$ which is not written for simplicity). Figure shows schematically the probability density (\ref{Wigner_XY-space}) evolving in time after $N$ periods as a circular rotation (here $\eta_y/\eta_x=1$).}
\label{Fig-Results}
\end{figure}

To illustrate the meaning of the obtained result (\ref{Psi(t)}), we calculate the corresponding Wigner function of the mechanical subsystem, based on the reduced density matrix $\hat \rho_m(t)=\text{Tr}_q \hat\rho (t)$, where $\hat \rho(t)=\vert \mathbf \Psi(t)\rangle\langle\mathbf \Psi(t)\vert$ is complete density matrix of the system, from which we trace out the charge-qubit degrees of freedom.
The Wigner function, defined as
%
%%%%%%%%%%%%%%%%%%%%%%%%
\begin{eqnarray}\label{Wigner_def}
&W&(x,p_x;y,p_y\vert t) = \frac{1}{(2\pi\hbar)^2}\int d\xi_x d\xi_y e^{-\tfrac{i}{\hbar} \left(p_x \xi_x+\imath p_y \xi_y\right)} \nonumber\\
&\times & \langle x+\tfrac{\xi_x}{2}\vert\otimes\langle
y+\tfrac{\xi_y}{2}\vert \,\, \hat\rho_m(t) \,\, \vert x-\tfrac{\xi_x}{2}\rangle\otimes\vert y-\tfrac{\xi_y}{2}\rangle, \,\,\,\,\,\, 
\end{eqnarray}
%%%%%%%%%%%%%%%%%%%%%%%%
%
reads (for simplicity we set $\hbar = 1$)
%
%%%%%%%%%%%%%%%%%%%%%%%%
\begin{eqnarray}\label{Wigner}
&&\hspace{-0.5cm}W(x,p_x;y,p_y\vert t)= \nonumber\\
&&\hspace{0.5cm}\frac{\vert c_+\vert^2}{\pi^2}
\exp\left[-\left(x-\mathcal{R}_x(\tau)\right)^2-\left(p_x-\mathcal{P}_x(\tau)\right)^2
\right. \nonumber\\
&&\hspace{2.1cm}- \left. \left(y-\mathcal{R}_y(\tau)\right)^2 -\left(p_y-\mathcal{P}_y(\tau)\right)^2\right] \nonumber\\
&&\hspace{0.2cm}+\frac{\vert c_-\vert^2}{\pi^2}
\exp\left[-\left(x+\mathcal{R}_x(\tau)\right)^2-\left(p_x+\mathcal{P}_x(\tau)\right)^2 \right. \nonumber\\
&&\hspace{2.1cm}-\left. \left(y+\mathcal{R}_y(\tau)\right)^2-\left(p_y+\mathcal{P}_y(\tau)\right)^2\right],
\end{eqnarray}
%%%%%%%%%%%%%%%%%%%%%%%%
%
where
%
%%%%%%%%%%%%%%%%%%%%%%%%
\begin{eqnarray}\label{Shifts}
\mathcal{R}_x(\tau) &=& \eta_x N\cos\omega\tau, \nonumber\\
\mathcal{P}_x(\tau) &=& -\eta_x N\sin\omega\tau, \nonumber\\
\mathcal{R}_y(\tau) &=& -\eta_y N\sin\omega\tau, \nonumber\\
\mathcal{P}_y(\tau) &=& -\eta_y N\cos\omega\tau
\end{eqnarray}
%%%%%%%%%%%%%%%%%%%%%%%%
%
describe the time-dependent shifts from the origin in real ($\mathcal R$) and momentum ($\mathcal P$) space. Integrating out the momenta, we project the Wigner function onto the $(x,y)$-plane, obtaining the probability density of the mechanical subsystem evolving in time (after $N$ periods, i.e. $t>NT$, $\tau=t-NT$)
%
%%%%%%%%%%%%%%%%%%%%%%%%
\begin{eqnarray}\label{Wigner_XY-space}
W(x,y\vert t)&=& \frac{\vert c_+\vert^2}{\pi}
\exp\left[-\left(x-\mathcal{R}_x(\tau)\right)^2-\left(y-\mathcal{R}_y(\tau)\right)^2 \right] \nonumber\\
&+&\frac{\vert c_-\vert^2}{\pi}
\exp\left[-\left(x+\mathcal{R}_x(\tau)\right)^2-\left(y+\mathcal{R}_y(\tau)\right)^2 \right]. \nonumber\\
\end{eqnarray}
%%%%%%%%%%%%%%%%%%%%%%%%
%
Result is shown in Fig. \ref{Fig-Results}, clearly presenting mechanical rotation in the plane between electrodes being the goal of applied time-evolution protocol. Analysis of the time-dependent shifts (\ref{Shifts}) indicates rotation along an ellipse with ratio of half-axes determined by the ratio $\eta_y/\eta_x$ (if it is 1, motion is along the circle).

%
%
%%%%%%%%%%%%%%%%%%%%%%%%%%
%%%%%%%%%%%%%%%%%%%%%%%%%%%%
%
%

\section{Conclusions}

In this paper we suggest a NEM setup and time-evolution protocol to operate external parameters, i.e. voltages applied to electrodes, to achieve controlled mechanical in-plane rotation of the charge qubit. The setup operates in the AC Josephson regime, comprising the voltage-biased superconducting electrodes, gate electrodes and charge qubit (superconducting grain in the regime of the Cooper pair box) placed on the top of suspended nanomechanical cantilever performing oscillations in the $(x,y)$-plane. Resonant tunneling of Cooper pairs between the charge qubit and superconducting electrodes (Josephson frequency is equal to mechanical frequency) builds nanomechanical vibrational coherent states, which, under the action of electric field provided by voltage at the gate electrodes, develop into the two-dimensional cat states entangled with charge qubit states. By the particular choice of high frequency gate voltage (with frequency equal to double mechanical frequency) the specific type of coherent states is built over time: as shown by the time-evolving probability density of mechanical subsystem projected on the real space (the $(x,y)$-plane), the time-dependent resulting wave presents circularly moving 2D coherent states entangled with qubit states, which in essence describes designed rotating motion of the superconducting grain. In general, motion is of elliptical shape determined by the coupling parameters to the superconducting and gate electrodes.

The physical feasibility of the proposed setup was mainly discussed in our previous paper on this topic \cite{npjDR} performing number of numerical calculations simulating different decoherence and dephasing processes, mismatch from the resonance condition, control of the applied external voltages etc. Fabrication process of creating the nanopillars (vibrating cantilever - see for example Ref. \cite{CKim}) are heading towards 1GHz operating nanomechanical frequencies with huge quality factors $Q \ge 10^5$. The zero-point amplitude of motion is then of the order of 1-10 pm, tunneling length of the order of 1 - 10 \AA, while the size of the tunneling contacts on terminals is of the order of 10 - 100 nm, within the reach of modern e-beam lithographic techniques.
Bias voltages of the order of $10 \,\, \mu$V is controllable down to $0.1 \% $ (e.g. by Keysight B2961A) which should satisfy the requirements.

As shown in this paper, the two-dimensional nanomechanical cat states, moving in space along circular trajectory while being entangled with superconducting charge qubit, can be generated.
The motivation lays in rather ambitious vision to achieve quantum communication through controlled spatial motion of qubits, carrying the quantum information, between nodes of the quantum network. It is the concept of so-called "flying qubits".
This research opens an intriguing possibility to design, so to say, a "quantum wiring", i.e, having in mind properties of the quantum-mechanical wave function that contains superposition of entangled objects at different positions, to create a quantum trajectory for quantum communication. It is achieved without external nanomechanical driving, but controlling all functionality electrically instead.
The nanomechanical implementation of the latter is motivated by its compactness and amazingly high quality factors achievable nowadays in nanomechanics, certainly crucial for any process involving quantum information. 
Further research is required to clarify how far the design of quantum trajectory facilitated by nanomechanical cat states demonstrated above can be utilized for the purpose of quantum communication.
%This includes investigation of various time-dependent operating protocols on gate electrodes, to induce different types of coupling and corresponding types of controlled nanomechanical motion.

\section*{Acknowledgements}

This work was supported by the QuantiXLie Centre of Excellence, a project co-financed by the Croatian
Government and European Union through the European Regional Development Fund - the Competitiveness and Cohesion Operational Programme (Grant PK.1.1.02), and by IBS-R024-D1. The authors are grateful to the PCS IBS, Daejeon, Republic of Korea for the hospitality while working on this paper.

\end{document}